\documentclass{aa} 
\usepackage{graphicx} 

\begin{document}

   \title{Abundance determinations in H\,{\sc ii} regions}

   \subtitle{Model fitting versus T$_e$ -- method}

\author{L.S.~Pilyugin \inst{1}}

  \offprints{L.S. Pilyugin }

   \institute{   Main Astronomical Observatory
                 of National Academy of Sciences of Ukraine,
                 27 Zabolotnogo str., 03680 Kiev, Ukraine, 
                 (pilyugin@mao.kiev.ua)
                 }
                 
\date{Received 30 September 2002 / accepted 18 October 2002}

\abstract{
The discrepancy between the oxygen abundances in high-metallicity H\,{\sc ii} 
regions determined through the T$_e$ -- method (and/or through the corresponding 
"strong lines -- oxygen abundance" calibration) and that determined through the 
model fitting (and/or through the corresponding "strong lines -- oxygen 
abundance" calibration) is discussed. It is suggested to use the interstellar 
oxygen abundance in the solar vicinity, derived with very high precision from 
the high-resolution observations of the weak interstellar OI$\lambda$1356 
absorption lines towards the stars, as a "Rosetta stone" to verify the validity 
of the oxygen abundances derived in H\,{\sc ii} regions with the T$_e$ -- method 
at high abundances. The agreement between the value of the oxygen abundance at 
the solar galactocentric distance traced by the abundances derived in H\,{\sc ii} 
regions through the T$_e$ -- method and that derived from the interstellar 
absorption lines towards the stars is strong evidence in favor of that 
i) the two-zone model for T$_e$ seems to be a realistic interpretation of the 
temperature structure within H\,{\sc ii} regions, and 
ii) the classic T$_e$ -- method provides accurate oxygen abundances in 
H\,{\sc ii} regions. 
It has been concluded that the "strong lines -- oxygen abundance" 
calibrations must be based on the H\,{\sc ii} regions with the oxygen abundances 
derived with the T$_e$ -- method but not on the existing grids of the models 
for H\,{\sc ii} regions.
}             

\titlerunning{Abundance determinations in H\,{\sc ii} regions: model fitting 
versus T$_e$ -- method}

\authorrunning{L.S.~Pilyugin}

\maketitle

\section{Introduction}

An investigation of variations of chemical properties among galaxies is very 
important for the development of the theory of the structure and evolution of 
galaxies. Accurate abundances are necessary for such investigations. 
Good spectrophotometry of H\,{\sc ii} regions is now available for a large 
number of galaxies, and the realibility of abundances is mainly defined by the 
method for abundance determination in H\,{\sc ii} regions.

Abundance in H\,{\sc ii} regions can be derived from measurements of temperature-sensitive 
line ratios, such as [OIII]$\lambda \lambda$4959,5007/[OIII]$\lambda$4363. 
Following Stasinska (2002b), this classical T$_{e}$ - method will be referred to 
as the direct empirical method. The abundance in H\,{\sc ii} regions can be also 
derived through photoionization model fitting. This method for abundance 
determination will be referred to as the theoretical (or model) method.

Unfortunately, in oxygen-rich H\,{\sc ii} regions the temperature-sensitive lines such 
as [OIII]$\lambda$4363 are often too weak to be detected. For such H\,{\sc ii} regions, 
abundance indicators based on more readily observable lines were suggested 
(Pagel et al. 1979; Alloin et al. 1979). The oxygen abundance indicator 
R$_{23}$ = ([OII]$\lambda \lambda$3727,3729 + [OIII]$\lambda \lambda$4959,5007)/H$_{\beta}$, 
suggested by Pagel et al. (1979), has found widespread acceptance and use for the 
oxygen abundance determination in H\,{\sc ii} regions where the temperature-sensitive 
lines are undetectable. 
The strategy of this way of abundance determination is very simple: the relation 
between strong oxygen line intensities and oxygen abundances is established 
based on the H\,{\sc ii} regions in which the oxygen abundances are determined through 
the T$_e$ -- method, and then this relation is used for the abundance determination 
in H\,{\sc ii} regions in which the temperature-sensitive lines are not available. 
The relation (between strong oxygen line intensities and oxygen abundances) 
established on the basis of H\,{\sc ii} regions in which the oxygen abundances are 
determined through the T$_e$ -- method (direct empirical method) will be 
referred to as empirical calibration. 

The grids of photoionization models 
are often used to establish the relation between strong oxygen line intensities 
and oxygen abundances (Edmunds \& Pagel 1984, McCall et al. 1985, Dopita \& 
Evans 1986, Kobulnicky et al 1999, Kewley \& Dopita 2002, among others).
The relation (between strong oxygen line intensities and oxygen abundances) 
established on the basis of the grids of the photoionization models for 
H\,{\sc ii} regions will be referred to as theoretical or model calibration. 
 
The early calibrations were one-dimensional (Edmunds \& Pagel 1984, McCall 
et al. 1985, Dopita \& Evans 1986, Zaritsky et al. 1994), i.e. the relation of 
the type $O/H = f(R_{23})$ was used. It has been shown (Pilyugin 2000, 2001a,b) 
that the error in the oxygen abundance derived with the one-dimensional 
calibrations involves a systematic error. The origin of this systematic error 
is evident. In a general case, the intensities of oxygen emission lines in 
spectra of H\,{\sc ii} region depend not only on the oxygen abundance but also on the 
physical conditions (hardness of the ionizing radiation and geometrical factors). 
Then in determininig the oxygen abundance from line intensities the 
physical conditions in the H\,{\sc ii} region should be taken into account. In the T$_e$ -- 
method this is done via T$_{e}$. In one-dimensional calibrations the physical 
conditions in H\,{\sc ii} regions are ignored. Starting from the idea of McGaugh (1991) 
that the strong oxygen lines contain the necessary information to determine 
accurate abundances in (low-metallicity) H\,{\sc ii} regions, it has been shown 
(Pilyugin 2000, 2001a,b) that the physical conditions in H\,{\sc ii} regions can be 
estimated and taken into account via the excitation parameter P. A two-dimensional or 
parametric calibration (P -- method) has been suggested. A more general relation 
of the type $O/H = f(P, R_{23})$ is used in the P -- method, compared to the 
relation of the type $O/H = f(R_{23})$ used in one-dimensional calibrations. 

It should be stressed that "strong lines -- oxygen abundance" calibrations do not 
form a uniform family. One should clearly recognize that there are two different 
types of calibrations. The calibrations of the first type are the empirical 
calibrations (established on the basis of H\,{\sc ii} regions in which the 
oxygen abundances are determined through the T$_e$ -- method). Two-dimensional 
empirical calibrations both at low and at high metallicities were recently 
derived by Pilyugin (2000, 2001a,c). The calibrations of the second type are 
the theoretical (or model) calibrations (established on the basis of the grids 
of the photoionization models of H\,{\sc ii} regions). The two-dimensional 
theoretical calibrations were recently suggested by Kobulnicky et al. (1999) 
and Kewley \& Dopita (2002).

Thus, at the present day there actually exist two scales of oxygen abundances 
in H\,{\sc ii} regions. The first (empirical) scale corresponds to the oxygen abundances
derived with the T$_e$ -- method or with empirical calibrations (the P -- 
method). The second (theoretical or model) scale corresponds to the oxygen 
abundances derived through the model fitting or with theoretical (model) 
calibrations. The comparison of these scales of oxygen abundances in H\,{\sc ii} regions 
and their evaluation are the goals of the present study.

\section{Oxygen abundances in H\,{\sc ii} regions: 
         model fitting versus T$_e$ -- method}

\begin{figure}
\resizebox{\hsize}{!}{\includegraphics[angle=0]{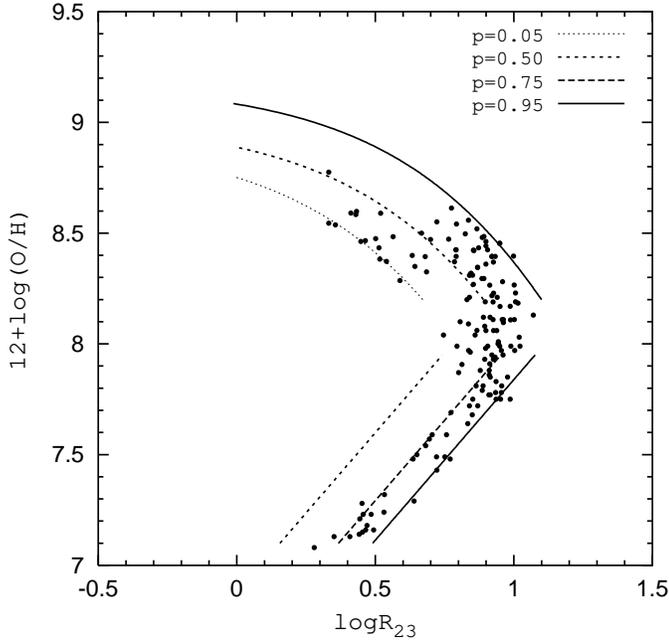}}
\caption{
The two-dimensional empirical calibration 
obtained by Pilyugin (2000, 2001c) (low-metallicity range) and Pilyugin (2001a) 
(high-metallicity range). Each O/H -- R$_{23}$ relation is labeled with 
corresponding value of the excitation parameter $P$. 
The points are H\,{\sc ii} regions with oxygen 
abundances determined through the T$_e$ -- method (compilation of data from 
Pilyugin 2000, 2001a).
}
\label{figure:pilyugin}                                     
   \end{figure}

Fig.\ref{figure:pilyugin} shows the two-dimensional empirical calibration 
obtained by Pilyugin (2000, 2001c) (low-metallicity range) and Pilyugin (2001a) 
(high-metallicity range). Each O/H -- R$_{23}$ relation is labeled with 
corresponding value of the excitation parameter $P$ which is defined as 
\begin{equation}
P =  \frac{[OIII] \lambda \lambda 4959, 5007}
              {[OII] \lambda 3727 + [OIII] \lambda \lambda 4959, 5007} .
\label{equation:p}
\end{equation}
The points in Fig.\ref{figure:pilyugin} are H\,{\sc ii} regions with oxygen 
abundances determined through the T$_e$ -- method (compilation of data from 
Pilyugin 2000, 2001a).

\begin{figure}
\resizebox{\hsize}{!}{\includegraphics[angle=0]{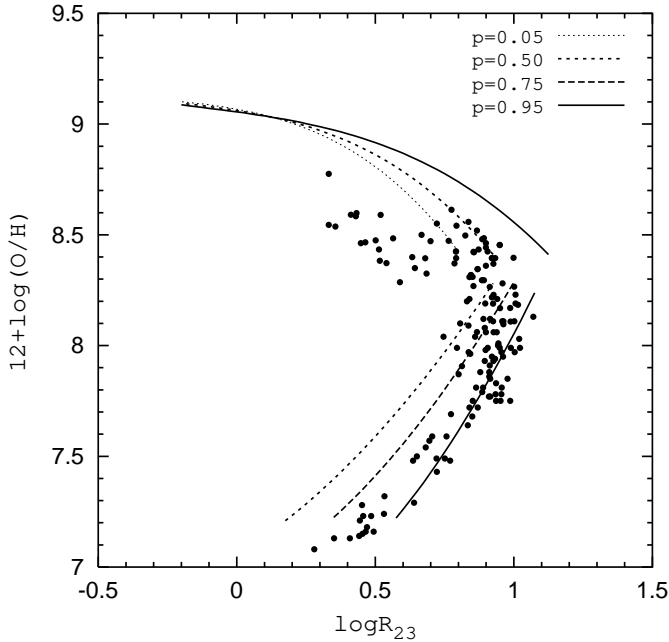}}
\caption{
The two-dimensional theoretical calibration reported by Kobulnicky et al. 
(1999). The lines are the O/H -- R$_{23}$ relations for different values of 
the excitation parameter P. Every line is labeled with corresponding value of 
the excitation parameter P. 
The points are the same data as in Fig.\ref{figure:pilyugin}. 
}
\label{figure:kobuln}                                     
   \end{figure}

Fig.\ref{figure:kobuln} shows the two-dimensional theoretical calibration 
reported by Kobulnicky et al. (1999). This calibration is  
based on the grid of the models of H\,{\sc ii} regions after McGaugh (1991).
The $O/H = f(R_{23})$ relations after Kobulnicky et al. (1999) are shown in 
Fig.\ref{figure:kobuln} with lines. Every line is labeled with corresponding value 
of the excitation parameter $P$. 
The ionization parameter $y$ was used in the two-dimensional theoretical 
calibration reported by Kobulnicky et al. (1999). 
The ionization parameter $y$ was defined as 
\begin{equation}
y = \log \frac{[OIII] \lambda \lambda 4959, 5007}
              {[OII] \lambda 3727} .
\label{equation:y}
\end{equation}
As can be seen from Eq.(\ref{equation:p}) and Eq.(\ref{equation:y}), 
the excitation parameter $P$ used in the two-dimensional empirical calibration 
of Pilyugin (2000, 2001a,c) and the ionization parameter $y$ used in the 
two-dimensional theoretical calibration reported by Kobulnicky et al. (1999)
are connected by a simple relation 
\begin{equation}
y = \log \frac{P} {1-P} .
\label{equation:yp}
\end{equation}

McGaugh (1991) has concluded that his grid of H\,{\sc ii} region models 
agrees with existing oxygen abundances determined through the T$_e$ -- 
method at low metallicities (his Figure 14), the rms of the residuals between 
his abundances and published abundances based on temperature-sensitive line 
ratios is 0.05dex with no zero-point offset. Inspection of 
Fig.\ref{figure:kobuln} confirms that the theoretical calibration 
of Kobulnicky et al. (1999), based on McGaugh's grid of  H\,{\sc ii} region 
models, agrees quantitatively with the recent oxygen abundances derived 
through the direct empirical method (the T$_e$ -- method) at low metallicities.
A comparison of Fig.\ref{figure:pilyugin} with Fig.\ref{figure:kobuln} shows 
that the discrepancy between the theoretical calibration of Kobulnicky et al. and 
the empirical calibration of Pilyugin is negligible small for the very 
low-metallicity (12+logO/H around 7.3), high-excitation (P around 0.95) 
H\,{\sc ii} regions, but the discrepancy increases with increasing 
metallicity and with decreasing excitation parameter, reaching the value of 
$\Delta$logO/H around 0.15 dex for H\,{\sc ii} regions with 
12+logO/H around 7.9. 
The agreement between the theoretical calibration of Kobulnicky et al. and 
the empirical calibration of Pilyugin disappears for H\,{\sc ii} regions that  
lie on the upper branch of the O/H -- R$_{23}$ diagram.

It should be noted that the agreement (at least quantitative) between the 
theoretical calibrations and the empirical calibrations at low metallicities is 
not a general rule. For example, there is a significant discrepancy between 
two-dimensional theoretical calibration of Kewley \& Dopita (2002) and the 
empirical calibration of Pilyugin at high and at low metallicities. 
The theoretical calibration of Kewley \& Dopita is also in 
conflict with the theoretical calibration of Kobulnicky et al.

Thus, the theoretical calibration of Kobulnicky et al. (1999) agrees 
quantitatively with the empirical calibrations of Pilyugin (2000, 2001a,b) 
at low metallicities and these calibrations are in conflict at high 
metallicities.

\section{Verification of the validity of the T$_e$ -- method at high metallicities} 

The validity of the T$_e$ -- method at high metallicities has been questioned 
in a number of investigations.
According to Stasinska (2002a,b), at high metallicities large temperature 
gradients are expected in ionized nebulae. Therefore, the T$_e$ -- method based 
on [OIII]$\lambda$4363/5007 will underestimate the abundances of heavy elements, 
since the  [OIII]$\lambda$4363 line will be essentially emitted in the high 
temperature zones, inducing a strong overestimate  of the average electron 
temperature. Therefore, although with very large telescopes it will now be 
possible to measure [OIII]$\lambda$4363 even in high metallicity giant H\,{\sc ii} 
regions, one should refrain from interpreting this line in the usual way. Doing 
this, one would necessary find sub-solar oxygen abundances, even for giant 
H\,{\sc ii} regions with metallicities well above solar. 

Thus, the validity of the T$_e$ -- method has been verified by comparison with 
the H\,{\sc ii} region models. As it can be seen in the previous section,  
the recent H\,{\sc ii} region models are not indisputable even at low 
metallicities. Why should one expect that H\,{\sc ii} region models provide 
more realistic abundances compared 
to the T$_e$ -- method at high metallicities? Indeed, according to Stasinska 
(2002b) this would be true if the constraints were sufficiently numerous (not
only on emission line ratios, but also on the stellar content and on the 
nebular gas distribution) and if the model fit were perfect (with a 
photoionization code treating correctly all the relevant physical processes and 
using accurate atomic data). These conditions are never met in practice. 
Abundances are not necessary better determined from model fitting.

Then, the validity of the T$_e$ -- method at high metallicities cannot be 
indisputably confirmed or rejected by comparison with the recent 
H\,{\sc ii} region models. Fortunately, there is another way to verify the 
validity of the T$_e$ -- method at high metallicities.

High-resolution observations of the weak interstellar OI$\lambda$1356 absorption 
lines towards the stars allow one to determine interstellar oxygen 
abundance in the solar vicinity with very high precision. It should be noted 
that this method is in fact model-independent.  These observations 
yield a mean interstellar oxygen abundance of 319 O atoms per 
10$^6$ H atoms (or 12+log(O/H) = 8.50) (Meyer et al. 1998, Sofia \& Meyer 2001).  
There are no statistically significant variations in the measured oxygen 
abundances from line of sight to line of sight; the rms scatter value for these
oxygen abundances is low, $\pm$0.05dex. Out to 1.5 kpc, the oxygen 
abundances are stable in diffuse clouds with different physical conditions as 
measured by the fraction of H in the form of H$_2$.   

Caplan et al (2000) and Deharveng et al (2000) have analysed Galactic 
H\,{\sc ii} regions and have obtained the slope --0.0395 dex/kpc with central 
oxygen abundance 12+log(O/H) = 8.82 and 12+log(O/H) = 8.48 at the solar 
galactocentric distance. 
All the available spectra of Galactic H\,{\sc ii} regions with measured 
[OIII]$\lambda$4363 lines were compiled by Pilyugin et al (2002b), and oxygen 
abundances in Galactic H\,{\sc ii} regions were recomputed in the same way,
using the T$_e$ -- method. These data result in 
oxygen abundance 12+log(O/H) = 8.50 at the solar galactocentric distance 
although the dispersion in derived abundances is relatively large. 
Thus, the value of the oxygen abundance at the solar galactocentric distance 
derived from consideration of the H\,{\sc ii} regions is in agreement with that 
derived with high precision from the interstellar absorption lines towards the 
stars. The agreement between the value of the oxygen 
abundance at the solar galactocentric distance traced by the abundances derived 
in H\,{\sc ii} regions through the T$_e$ -- method and that derived from the 
interstellar absorption lines towards the stars is strong evidence in favor of 
that 
{\it i)} the two-zone model for T$_e$ seems to be a realistic interpretation of the 
temperature structure within H\,{\sc ii} regions, and 
{\it ii)} the classic T$_e$ -- method provides accurate oxygen abundances in 
H\,{\sc ii} regions up to oxygen abundances as large as 12+log(O/H) = 8.60$\div$8.70. 
Thus, one can concluded that the H\,{\sc ii} regions with T$_e$-abundances provide 
a more reliable basis for calibration than the H\,{\sc ii} region models.

\begin{figure}
\resizebox{\hsize}{!}{\includegraphics[angle=0]{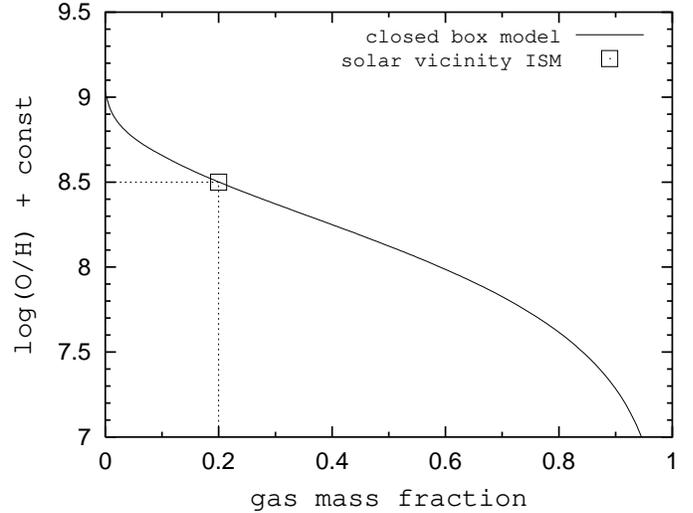}}
\caption{
The oxygen abundance versus gas mass fraction diagram. The constant was 
chosen in such a way that the value of the gas mass fraction $\mu$=0.2 
corresponds to the oxygen abundance as large as 12+logO/H=8.5. This comes from 
the data for the solar vicinity interstellar medium (see text).
}
\label{figure:muoh}                                     
   \end{figure}

Unfortunately, there are no H\,{\sc ii} regions with measured [OIII]$\lambda$4363 
lines at highest metallicities (or at lowest logR$_{23}$, Fig.\ref{figure:pilyugin}).
There are only two high-metallicity, 12+log(O/H) $\sim$ 8.9, H\,{\sc ii} regions with 
measured temperature-sensitive lines, see Kinkel \& Rosa (1994),
Castellanos et al. (2002). 
Therefore, we have to use an extrapolation of the calibration to derive the 
abundances in H\,{\sc ii} regions with lowest value of logR$_{23}$. The oxygen 
abundances in those H\,{\sc ii} regions predicted by the empirical calibration 
are significantly 
lower compared to abundances predicted by theoretical (model) calibrations. 
There is hovewer indirect way to test the 
reality of these predictions. The present-day oxygen abundance in the 
solar vicinity is 12+log(O/H)=8.50, and for the present-day gas mass fraction, 
0.15$\div$0.20 appears to be a reasonable value (Malinie et al. 1993).
Fig.\ref{figure:muoh} shows the O/H -- $\mu$ diagram, where $\mu$ is the 
gas mass fraction. The prediction of the closed-box model for the chemical 
evolution of galaxies is presented by the solid line in Fig.\ref{figure:muoh}. 
The constant for O/H values was chosen in such manner that the value of the gas 
mass fraction $\mu$=0.2 corresponds to the oxygen abundance as large as 
12+logO/H=8.5 as in the solar vicinity (square in 
Fig.\ref{figure:muoh}). Inspection of Fig.\ref{figure:muoh} shows that there is 
no the possibility of large central (intersect) oxygen abundances (as large as 
12+logO/H $\sim$ 9.50 and higher) derived with theoretical (model) calibrations
(Zaritsky et al 1994, Garnett et al 1997). On the contrary, the low central oxygen 
abundances predicted by the empirical calibration (Pilyugin et al. 2002a) 
fit this picture well. 

It has been known for a long time that permitted lines in H\,{\sc ii} regions 
indicate higher oxygen abundances than forbidden lines. Peimbert (1967) 
suggested that the presence of spatial temperature fluctuations in gaseous 
nebulae can significantly influence the oxygen abundance in H\,{\sc ii} 
regions derived from forbidden lines. In contrast, permitted lines are almost 
independent of such variations and, in principle, they should be more precise 
indicators of the true chemical abundances. Esteban et al. (1998) found oxygen 
abundances for two positions in the Orion nebula from permitted and forbidden 
lines. The oxygen abundances derived from forbidden lines are coincident for 
both positions in the Orion nebula (12 + logO/H = 8.47) and agree well with the 
interstellar oxygen abundance in the solar vicinity derived from interstellar 
absorption lines towards the stars (12 + logO/H = 8.50). They found from 
permitted lines the oxygen abundance 12 + logO/H = 8.61 for position 1 and 
12 + logO/H = 8.68 for position 2, although the abundances obtained from 
the different multiplets observed show significant dispersion. 
The gas-phase oxygen abundance 12 + logO/H = 8.64 and the total (gas+dust) 
oxygen abundance as large as 12+logO/H = 8.72 were proposed by Esteban et al. 
(1998) for the Orion nebula. If permitted lines indicate the true oxygen abundance 
in the Orion nebula, then the uncertainty around 0.1 dex in the oxygen abundances 
derived from forbidden lines cannot be excluded. However, the origin of the 
discrepancy between abundances derived from permitted and forbidden lines is 
not indisputable (see discussion in Stasinska, 2002b). If the total (gas+dust) 
oxygen abundance in the Orion nebula coincides with the total oxygen abundance 
in the interstellar medium in the solar vicinity, and if permitted lines provide 
true gas-phase oxygen abundance in the Orion nebula, then the difference between 
gas-phase oxygen abundance in the Orion nebula and in the interstellar gas 
suggests that the depletion of oxygen into dust grains in the interstellar 
medium is around 0.1 dex higher than in the Orion nebula (and absolute depletion 
of oxygen in the interstellar medium is around 0.2 dex). It can be verified,
in principle, by measurement of the abundance of the same noble gas in the 
interstellar medium and in the Orion nebula. Unfortunately, measurements of 
the abundance of only the noble gas krypton in the interstellar medium are 
available (Cardelli \& Meyer 1997), but there is no data on the krypton abundance 
in the Orion nebula.
 
Thus it appears that the classic T$_e$ -- method provides more accurate oxygen 
abundances in H\,{\sc ii} regions at high metallicities as compared to the 
model fitting, although the uncertainty around 0.1dex cannot be excluded. 
As a consequence, the empirical calibration appears to be more 
justified than the theoretical (model) calibration.

\section{Conclusions}

The oxygen abundances in H\,{\sc ii} regions determined through the direct 
empirical method (the classic T$_e$ -- method) and/or through the corresponding 
empirical "strong lines -- oxygen abundance" calibration are compared with 
abundances determined through the model fitting and/or through the corresponding 
theoretical (model) "strong lines -- oxygen abundance" calibration. 
It was shown that the theoretical calibration of Kobulnicky et al. (1999) agrees 
quantitatively with the empirical calibrations of Pilyugin (2000, 2001a,b) 
at low metallicities and these calibrations are in conflict at high 
metallicities.

It is suggested to use the interstellar oxygen abundance in the solar 
vicinity, derived with very high precision from the high-resolution observations 
of the weak interstellar OI$\lambda$1356 absorption line towards the stars, as 
a "Rosetta stone" to verify the validity of the oxygen abundances derived in 
high-metallicity H\,{\sc ii} regions with the T$_e$ -- method. 
The agreement between the value of the oxygen 
abundance at the solar galactocentric distance traced by the abundances derived 
in H\,{\sc ii} regions through the T$_e$ -- method and that derived from the 
interstellar absorption lines towards the stars is strong evidence in favor 
of that 
i) the two-zone model for T$_e$ seems to be a realistic interpretation of the 
temperature structure within H\,{\sc ii} regions, and 
ii) the classic T$_e$ -- method provides accurate oxygen abundances in 
H\,{\sc ii} regions, although the uncertainty around 0.1dex cannot be excluded.

It has been concluded that at high metallicities the "strong lines -- oxygen 
abundance" calibrations must be based on the H\,{\sc ii} regions with the 
oxygen abundances derived through the T$_e$ -- method but not on the existing 
grids of the models for H\,{\sc ii} regions.

\begin{acknowledgements}
I thank B.E.J.~Pagel, G.~Stasinska, J.M.~Vilchez, L.J.~Kewley for useful 
discussions. I thank the anonymous referee for constructive comments. 
This study was partly supported by the Joint Research Project between 
Eastern Europe and Switzerland (SCOPE) No. 7UKPJ62178, the NATO grant 
PST.CLG.976036, and the Italian national grant delivered by the MURST. 
\end{acknowledgements}

\end{document}